\documentclass[aps,showpacs,graphicx]{revtex4} 

\usepackage{amssymb} 
\usepackage[dvips]{graphicx} 
\usepackage[english]{babel} 
\usepackage{braket}
\newcommand{\beq}{\begin{equation}} 
\newcommand{\eeq}{\end{equation}} 
\newcommand{\beqn}{\begin{eqnarray}} 
\newcommand{\eeqn}{\end{eqnarray}} 
\newcommand{\bsigma}{\mbox{\boldmath $\sigma$}} 
\newcommand{\btau}{\mbox{\boldmath $\tau$}} 
\newcommand{\half}{\frac{1}{2}} 

\newcommand{\br}{{\bf r}}

\newcommand{\yspin}{{\cal Y}}
\newcommand{\ripm}{\rho^{\rm IPM}}
\newcommand{\rsrc}{\rho^{\rm SRC}}
\newcommand{\rlrc}{\rho^{\rm LRC}}
\usepackage[usenames]{color} 
\usepackage{ulem}

\begin{document} 
 
\noindent 
\title{Charge radii of Ca isotopes and correlations} 
 
\author{G. Co'$^{\,1,2}$, M. Anguiano$^{\,3}$, A. M. Lallena$^{\,3}$ }
\affiliation{$^1$ Dipartimento di Matematica e Fisica ``E. De Giorgi'', 
  Universit\`a del Salento, I-73100 Lecce, ITALY, \\ 
$^2$ INFN Sezione di Lecce, Via Arnesano, I-73100 Lecce, ITALY, \\ 
$^3$ Departamento de F\'\i sica At\'omica, Molecular y 
  Nuclear, Universidad de Granada, E-18071 Granada, SPAIN
}  

\date{\today} 

\bigskip 
 
\begin{abstract} 
We study the effects of short- and long-range correlations on the charge radii of 
Ca isotopes. We start our investigation with an independent particle model consisting 
in Hartree-Fock plus  Bardeen-Cooper-Schrieffer
calculations with finite-range effective nucleon-nucleon interactions of 
Gogny type. The short-range correlations effects are evaluated by considering 
all the terms of a cluster expansion containing a single correlation line. The long-range 
correlations are taken into account by including the coupling with the quasi-particle
random phase approximation phonons. While the effects of the short-range correlations
are negligible, those of the long-range correlations largely modify the independent 
particle model results and improve the agreement with the experimental data.
\end{abstract}

\bigskip 
\bigskip 
\bigskip 
 
%\pacs{21.60.Jz; 25.40.Kv} 

\maketitle 
\section{Introduction}

In these last few years, the measurement of several isotope shifts completed
the information about the charge radii of Ca nuclei.
Garcia Ruiz {\it et al.} \cite{gar16} measured the isotope shifts 
of the neutron rich $^{49,51,52}$Ca nuclei using laser spectroscopy 
and Miller {\it et al.} \cite{mil19} investigated the proton rich $^{36,37,38}$Ca 
isotopes with similar experimental techniques. These new data have been 
used to derive the values of the charge radii of these nuclei. 
 
The behavior of the experimental values of the charge radii shows 
a steep enhancement for nuclei with $A > 48$. This behaviour is not
described by several independent particle models  \cite{gar16,mil19,tan20}). 
It has been shown \cite{mil19} that by using elaborate energy functionals, 
containing a large number of free parameters \cite{fay00,klu09,rei17}, 
it is possible to reproduce the full set of experimental data. 
However, the physics simulated by these new functionals is difficult to disentangle.

In this work we attack the problem by using a more traditional independent particle
model, which in our case consists in Hartree-Fock (HF) plus 
Bardeen-Cooper-Schrieffer (BCS) calculations,  
and we study the effects of the correlations. We consider short-range correlations
related to the strongly repulsive core of the nucleon-nucleon interaction and long-range correlations
generated by the coupling of collective nuclear excitations to the single-particles degrees of freedom. 

This work is organized as follows. Sec. II is devoted to present some details of the 
various theoretical approaches used in our calculations. 
In Sec. III we show our results 
which are discussed in Sec. IV where we also present our conclusions.

\section{Theoretical approaches}

The goal of our study is the evaluation of the root mean square (r.m.s.) radii of nuclear density distributions
$\rho_\alpha(r)$:
\beq
R_\alpha\, \, = \, \left[ \displaystyle \frac 
{\displaystyle \int_0^\infty {\rm d}r \, r^4 \, \rho_\alpha (r)}
{\displaystyle \int_0^\infty {\rm d}r \, r^2 \, \rho_\alpha (r)} 
\right]^{\half}
\, .
\label{eq:radius}
\eeq
Proton and neutron r.m.s. radii are obtained by considering the (point-like) proton, $\rho_{\rm p}$, and neutron, $\rho_{\rm n}$, density distributions, which are defined as:
\beq
\rho_\alpha (\br)\, =\, \frac { A } {{\bigl\langle}{\Psi}|{\Psi} {\bigr\rangle} } \, {\Bigl\langle} {\Psi} \Bigl| \displaystyle {\sum_j}^\prime \, \delta(\br-\br_j) \Bigr| {\Psi} {\Bigr\rangle} \, , \,\,\, \alpha \, \equiv \, {\rm p, n} \, ,
\label{eq:rho}
\eeq
where $\ket{\Psi}$ is the state describing the full nucleus, which has $A$ nucleons ($Z$ protons and $N$ neutrons). 
In the above equation, and in all the following ones, $\sum'$ indicates that the sum is limited 
either to the protons or to the neutrons as indicated by the sub-index $\alpha$. 
Mass r.m.s. radii are obtained by using the mass density distributions, $\rho_{\rm m}=\rho_{\rm p}+\rho_{\rm n}$.
We calculate the charge r.m.s. radii by inserting in Eq.~(\ref{eq:radius}) the charge distributions, $\rho_{\rm c}$ obtained  by folding the (point-like) proton densities, $\rho_{\rm p}$, with the charge proton form factor. We used a dipole parameterization of this form factor \cite{pov93}, having verified that other, more complex, expressions produce differences smaller than the numerical accuracy of our calculations.

The aim of the various nuclear models we adopted in our study 
is the evaluation of the density distributions required to calculate the r.m.s radii.

\subsection{Independent particle model}

The starting point of our approach is the Independent Particle Model (IPM) where the nuclear state
$\ket{\Psi}$ is described as a Slater determinant, $\ket{\Phi}\equiv \Phi(1,2,\ldots,A)$, of single particle (s.p.) wave functions.
We exploit the spherical symmetry of the problem, and for the s.p. wave functions we have used the expression \cite{ang14,ang15}
\beq
{\phi_{\mu}}(\br)\, \equiv \,
{\phi_{nljm}^{m_t}}(\br)\,=\,{\cal R}_{nljm}(r) \, {\yspin_{ljm}}(\Omega) \, \chi_{\frac{1}{2}\,m_t} \, ,
\label{eq:philjm}
\eeq
where we have indicated with $n$ the principal quantum number, and with
$l$, $j$ and $m$ the quantum numbers identifying the orbital, the total angular momentum
and its projection on the $z$ axis, respectively. The symbol 
$\yspin$ indicates the spin-spherical harmonics \cite{edm57}, 
and $m_t$ the third component of the isospin.

Since we have considered nuclei without deformation, 
we have assumed a unique radial wave function for the same $n,l,j$
quantum numbers, i.e. ${\cal R}_{nljm} \equiv {\cal R}_{nlj}$. 
In  open shell nuclei the degenerate, not fully occupied, s.p. neutron state near the Fermi level is 
equally occupied with respect to the projection quantum number $m$.
 
The first step of our IPM calculation consisted in obtaining the s.p. radial wave 
functions ${\cal R}_{nlj}$ by solving a set of HF equations. 
In a second step, we have considered the pairing effects by using these HF s.p. wave functions 
in a BCS calculation wich modifies the occupation probability, $v_\mu^2$, of each s.p. state. 
We named this IPM type of calculation HF+BCS \cite{ang14,ang15}, and, in this model, 
the density distributions are given by:
\beq
\ripm_\alpha(r) \,=\, \displaystyle \frac{1}{4\pi} \, {\sum_{\mu}}^\prime \,(2j +1) \, v^2_{\mu} \, \left[ {\cal R}_{\mu}(r) \right]^2 \, , \,\,\, \alpha \, \equiv \, {\rm p, n} \, .
\eeq

The only physical input of this calculation is the effective nucleon-nucleon interaction. We  have
consistently used the same finite-range effective nucleon-nucleon interactions of Gogny type 
\cite{dec80} in both the HF and the BCS steps. 
In order to test the sensitivity of our results to the effective 
nucleon-nucleon interaction we carried out
calculations with three different parameterizations. Two of them are the well known 
D1S  \cite{ber91} and D1M  \cite{gor09} interactions, and  
the third one is the D1ST2a, a force built by adding tensor terms 
to the D1S \cite{ang12}.

\subsection{Short-range correlations}

In the IPM each nucleon is free to move independently of the presence of the other ones. 
On the other hand, the nucleon-nucleon interaction has a strongly repulsive core which
prevents two nucleons from approaching each other at distances smaller than about $0.5\,$fm.
This is the source of the Short-Range Correlations (SRC). 

We have evaluated the effects of the SRC by following the approach of Ref.~\cite{co95} where
the nuclear state is described as
\beq
|\Psi \rangle \, \equiv \, \Psi^{\rm SRC} (1,2,\ldots,A)\,=\, F(1,2,\ldots,A) \, \Phi (1,2,\ldots,A) \, ,
\eeq
with $\Phi$ indicating the IPM Slater determinant, and $F$ a many-body correlation function
which we have assumed to be of the form
\beq
F(1,2,\ldots,A)\, = \, {\cal S} \, {\prod_{i<j}} \, \sum_{p=1}^6 \, f^{(p)} (r_{ij}) \, {O_{i,j}^{(p)}} \, ,
\label{eq:fcorrel}
\eeq 
We indicate with ${\cal S}$ a symmetrization operator, 
with $f^{(p)}(r_{ij})$ a two-body correlation scalar function acting on the $(i,j)$
nucleon pair, and with $\{O^{(p)},\,p=1,\ldots,6\}$ 
a set of operators depending on spin, isospin and tensor terms,
classified as in the usual Urbana-Argonne sequence \cite{wir95}. 
Specifically, we have considered operators of central type
\beq
O_{ij}^{(1)} \, = \, 1 \, , \,\,\, 
O_{ij}^{(2)} \, = \btau(i) \cdot \btau(j) \, , \,\,\,
O_{ij}^{(3)} \, =\bsigma(i) \cdot \bsigma(j) \, , \,\,\,
O_{ij}^{(4)} \, =\bsigma(i) \cdot \bsigma(j)   \btau(i) \cdot \btau(j) \, , 
\eeq
and of tensor type
\beq
O_{ij}^{(5)} \,= S(i,j) \, , \,\,\,
O_{ij}^{(6)} \,=S(i,j)\;\btau(i) \cdot \btau(j)  \, ,
\eeq
where $\bsigma$ indicates the spin operator, $\btau$ the isospin operator, 
and $S(i,j)$ the tensor operator defined as:
\beq
S(i,j) \, = 3\, \frac{\left[ \bsigma(i) \cdot \br_{ij} \right] \,
\left[ \bsigma(j) \cdot \br_{ij} \right]}{\br^2_{ij}} \, - \, \bsigma(i) \cdot \bsigma(j) \, .
\eeq

\begin{figure}[!t] 
\begin{center} 
\includegraphics [scale=0.4,angle=0]{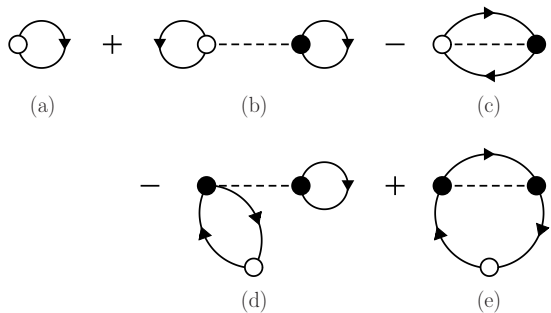} 
%\vskip -6.0 cm 
\caption{\small 
The set of SRC diagrams considered in our calculations. 
The dashed lines indicate the correlation function $h$, Eq. (\ref{eq:hdef}), 
while the oriented lines represent the IPM one-body density matrices $\ripm_\alpha(\br_1,\br_2)$, 
Eq. (\ref{eq:densmat}). 
The solid dots indicate a coordinate where an integration is carried out and the open ones those where the density is evaluated. 
}
\label{fig:diag1} 
\end{center} 
\end{figure}

The key point of the model of Ref. \cite{co95} is the truncation of the cluster expansion of the density 
distribution in order to consider only those terms with a single correlation function
\beq
h^{(p)}(r_{ij}) \, = \, f^{(p)}(r_{ij}) \,-\, \delta_{p,1}  \, ,
\label{eq:hdef}
\eeq
where $\delta$ is the Kronecker symbol. 
We show in Fig.~\ref{fig:diag1} the diagrams included in the calculations. The dashed lines represent the $h$ correlation
function and the oriented lines the IPM one-body density matrix
\beq
\ripm_\alpha(\br_1,\br_2) \,= \, {\sum_{\mu}}^\prime \, v^2_\mu \, \phi_\mu^*(\br_1) \, \phi_\mu(\br_2) \, , \,\,\, \alpha \, \equiv \, {\rm p, n} \, .
\label{eq:densmat}
\eeq
A solid dot indicates a point in $r$ space that is integrated, while the empty dot is the point 
where the density is calculated. 

Due to the orthonormality of the s.p. wave functions, the IPM density matrices 
satisfy the property
\beq
\int  {\rm d}^3 r^\prime \, \ripm_\alpha(\br_1,\br^\prime) \, \ripm_\alpha(\br^\prime,\br_2) \, =\,  \ripm_\alpha(\br_1,\br_2) \, .
\eeq
Because of this, the integration on the open dots in the diagrams of Fig. \ref{fig:diag1} implies that, 
in absolute value, the contributions of the diagrams (b) and (d) are equal, as well as those of 
the diagrams (c) and (e). For this reason, the integral on $\br$ of 
the correlated terms cancel exactly, and the correlated density is normalized as the IPM one:
\beq
\int  {\rm d}^3 r \, \rsrc_\alpha(\br) \, =\,  \int  {\rm d}^3 r \, \ripm_\alpha(\br) \, .
\eeq
These integrals are equal to $A$, $Z$ or $N$ depending on the density distribution considered. 
We indicate with $\rho(\br)$ the diagonal part of the one-body density matrix. 
It is worth remarking the need of considering all the diagrams of  Fig. \ref{fig:diag1}, i.e. those with two and
three points, in order to conserve the proper normalization of the correlated density. 

Explicit expressions of the contribution of the diagrams in terms of the radial s.p. wave functions are 
given in Ref. \cite{co95}. The only difference in the present calculations is the presence of the $v^2$ 
occupation probability multiplying every pair of s.p. wave functions.

\begin{figure}[!t] 
\begin{center} 
\includegraphics [scale=0.4,angle=0]{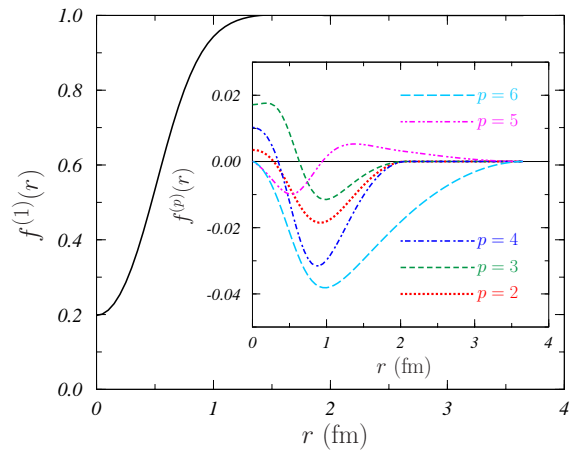} 
%\vskip -3.0 cm 
\caption{\small 
The SRC functions $f^{(1)}$ and $f^{({\rm p})}$ ($p=2,\ldots,6$), Eq. (\ref{eq:fcorrel}),
for the six operator channels considered in our calculations.
}
\label{fig:correl48} 
\end{center} 
\end{figure}

The two-body correlation scalar functions are the only physics input which is not consistently
derived from the selected effective nucleon-nucleon interaction.
We present here the results obtained by using the  $f^{({\rm p})}(r_{ij})$
taken from  a Fermi Hypernetted-Chain calculation for the $^{48}$Ca nucleus \cite{ari07}, carried out
with the microscopic V8' Urbana interaction \cite{wir95} (see Fig.~\ref{fig:correl48}).
We have tested the effects of other correlation
functions presented in Ref. \cite{ari07} related to other doubly magic nuclei and to nuclear matter, and
we did not find remarkable differences with the results presented here below. 

\subsection{Long-range correlations}

The part of the nuclear hamiltonian not considered by the IPM, the residual interaction, is the source of the
so-called Long-Range Correlations (LRC), which take into account collective phenomena inside the nucleus. 

We have described the effects of the LRC on the density distributions by extending the model presented in 
Refs.~\cite{row70,len90,ang01}. We have substituted the Random Phase Approximation $Y$ amplitudes 
with those obtained with the Quasi-particle Random Phase Approximation (QRPA) theory. 
In this approach, the density distributions can be written as:
\beq
\nonumber
\rlrc_\alpha(r) \, = \, \ripm_\alpha(r) \,- \, \sum_{J^\Pi} \, \frac{2J+1}{8\pi} \, \sum_{E_k} \, {\sum_{\mu < \nu}}^\prime \, \left| Y_{\mu \nu}^{J^\Pi} (E_k) \right|^2  \left\{
\left[ {\cal R}_\mu(r) \right]^2 \, - \, \left[ {\cal R}_\nu(r) \right]^2  \right\} \, , \,\,\, \alpha \, \equiv \, {\rm p, n} \, .
\label{eq:rholrc}
\eeq
From the above expression it emerges the fact that the relevant ingredient modifying the IPM density is provided
by the QRPA $Y$ amplitudes calculated for a specific excited state of energy $E_k$, 
angular momentum $J$ and parity $\Pi$. 
Because of the orthonormalization of the s.p. wave functions, 
an integration on $\br$ produces equal contributions of the two terms related to the same
QRPA $Y$ amplitudes, therefore they cancel with each other.
Also in this case the normalization of $\rlrc$ is the same as that of $\ripm$. 

Our QRPA calculations, whose technical details can be found in Ref.~\cite{don17}, are based on a 
discrete set of s.p. wave functions, therefore the energy spectrum we obtain is discrete. 
On the other hand, the three sums in Eq. (\ref{eq:rholrc}) have to be truncated. 
The dimensions of the s.p. configuration space, which limit
the sum on the states $\mu$ and $\nu$ and ensure the stability of the QRPA results,
are fixed by using the prescriptions described in Ref.~\cite{don17}. 

For a given multipolarity $J^\Pi$, we have studied the maximum energy required in the sum on $E_k$ 
to stabilize the result. Even though every excitation multipole, in each of the nuclei investigated, 
has its own value for this maximum energy $E_{\rm max}$, 
we have  found that $E_{\rm max} =20\,$MeV, for all multipolarities included 
in Eq.~(\ref{eq:rholrc}), ensures sufficient stability of our results. 

Concerning the sum on $J^\Pi$, we have considered all the positive and negative multipoles with angular momentum from $J=1$ to $J=3$ and also the contribution of $J=0^+$. As the more relevant contributions come from the low-lying excited states, in the case of the $^{42}$Ca, $^{44}$Ca and $^{46}$Ca isotopes, we have included 
also the $4^+$ and $6^+$ multipoles since, in our calculations, some excited state
with these multipolarities appears below $4\,$MeV.

%------------------

\section{Results}

In our study we have considered the even-even Ca isotopes. 
The parameters of the effective nucleon-nucleon interactions 
have been chosen to provide a good IPM description of 
the ground state of a set of nuclei all along the 
nuclear chart \cite{gor09,cha07t}.
As an example of  the performances of these interactions, 
we compare in Table~\ref{tab:be} the experimental binding 
energies  of the nuclei we have investigated \cite{bnlw}
with those obtained with our IPM (HF+BCS) by using the three 
interactions  considered. 
The good agreement between the empirical values and the results of our calculations is evident, 
and also expected, since binding energies, among other data, 
have been used to select the values of the force parameters
\cite{cha07t}.

\begin{table}[!ht]
\begin{center}
\begin{tabular}{ccccc}
\hline\hline
%& \multicolumn{4}{c}{Binding energies per nucleon (MeV)}\\
%\cline{2-5}
$A$ & D1S & D1ST2a & D1M & exp. \\
\hline
 34 & $7.254$ & $7.255$ & $7.148$ & $7.173$ \\
 36 & $7.867$ & $7.871$ & $7.754$ & $7.816$ \\
 38 & $8.296$ & $8.296$ & $8.180$ & $8.240$ \\
 40 & $8.626$ & $8.624$ & $8.513$ & $8.551$ \\
 42 & $8.668$ & $8.669$ & $8.556$ & $8.617$ \\
 44 & $8.711$ & $8.718$ & $8.602$ & $8.658$ \\
 46 & $8.719$ & $8.732$ & $8.614$ & $8.669$ \\
 48 & $8.694$ & $8.714$ & $8.593$ & $8.667$ \\
 50 & $8.551$ & $8.572$ & $8.458$ & $8.550$ \\
 52 & $8.404$ & $8.427$ & $8.320$ & $8.429$ \\
 54 & $8.211$ & $8.230$ & $8.140$ & $8.248$ \\
 56 & $8.028$ & $8.033$ & $7.956$ & $8.033$ \\
 58 & $7.832$ & $7.834$ & $7.768$ & $7.828$ \\
 60 & $7.614$ & $7.620$ & $7.577$ & $7.627$ \\
\hline \hline
\end{tabular}
\caption{Binding energies per nucleon, in MeV, for the Ca isotopes under investigation, 
obtained with our HF+BCS IPM by using the D1S, D1ST2a and D1M interactions, 
compared to the experimental values taken from the compilation 
of the Brookhaven National Laboratory \cite{bnlw}.
The largest experimental uncertainty is 0.012 MeV in $^{60}$Ca.
}
\label{tab:be}
\end{center}
\end{table}

Since our IPM is built to reproduce some experimental nuclear quantities, the appearance of phenomena
beyond IPM can be identified  by doing a relative comparison between the theoretical results. On the 
other hand, the quantity experimentally measured, the isotope shift, is defined in a relative manner
as the difference between the square of the charge radii of the each Ca isotope and that of $^{40}$Ca,
\beq
\delta R_{\rm ch}^2 (A) \,=\, \left[R_{\rm ch}(A)\right]^2 \, - \, 
\left[R_{\rm ch}(A=40)\right]^2 
\, .
\label{eq:shift}
\eeq
These are the two reasons that have induced us to concentrate our study on the isotope shifts rather than on
the charge radii.

\begin{figure}[!t] 
\begin{center} 
\includegraphics [scale=0.4,angle=0]{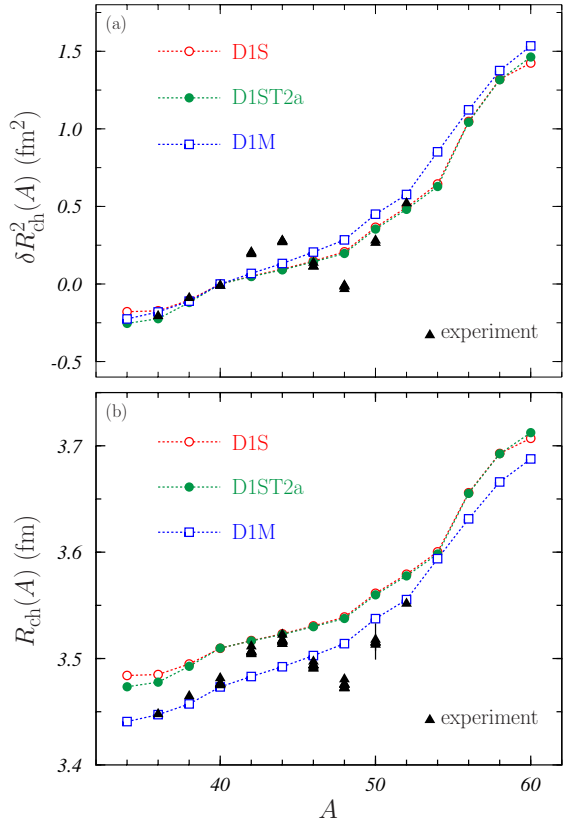} 
%\vskip -3.0 cm 
\caption{\small 
IPM results obtained with the D1S (open red dots), D1ST2a (solid green dots) 
and D1M (open blue squares) Gogny interactions for the Ca isotopes considered 
in our work, with the experimental data (black triangles). Panel (a): isotope shifts of the squared r.m.s. charge radii evaluated with respect to the
$^{40}$Ca r.m.s. radius, Eq. (\ref{eq:shift}); the experimental data have been obtained from Refs. \cite{gar16,mil19,woh81,pal84,ver92}.
Panel (b): r.m.s. charge radii of the Ca isotopes; the experimental values have been calculated by considering the experimental isotope shifts  shown in panel (a) and the reference r.m.s. charge radius $R_{\rm ch}(A=40)=3.4776(19)\,$fm \cite{ang13}. . 
}
\label{fig:IPM} 
\end{center} 
\end{figure}

In Fig.~\ref{fig:IPM}a, we compare the values of the isotope shifts
obtained in our IPM calculations, carried out 
with the three different parameterizations of the Gogny interaction, D1S (red open dots), 
D1M (blue open squares) and D1ST2a (green solid dots), 
with the available experimental data of Refs. \cite{gar16,mil19,woh81,pal84,ver92} (black triangles).

The first remark is that the behaviors of our IPM results are very similar, independently of the interaction used. 
A quantitative indicator of this similarity is the average of the absolute differences between these curves, which is smaller than $0.07\,$fm$^2$, with a maximum value of $0.2\,$fm$^2$ (in the case of $^{54}$Ca). 
It is also worth noting that 
the tensor terms of the force do not play any relevant role: the results obtained with the D1ST2a and D1S
almost overlap.

The comparison with the experimental data is not straightforward. 
We observe good agreement with the isotope shifts of the 
nuclei lighter than $^{40}$Ca and heavier than $^{48}$Ca, 
while in the cases of  $^{42}$Ca, $^{44}$Ca and $^{48}$Ca 
the IPM fails to describe the data. 
The IPM isotope shifts increase  smoothly with increasing neutron number. On the contrary, 
the experimental data show a steeper increase from 
$^{40}$Ca up to the value of $^{44}$Ca and a decreasing 
behavior for $^{46}$Ca and  $^{48}$Ca, whose isotope 
shift is almost zero.

A direct comparison between the r.m.s. radii obtained by using our IPM densities,
as indicated by Eq.~(\ref{eq:radius}), and the empirical ones is shown in Fig.~\ref{fig:IPM}b. 
The latter values have been obtained from Eq.~(\ref{eq:shift}) by considering 
the experimental isotope shifts and 
the reference value $R_{\rm ch}(A=40)=3.4776(19)\,$fm \cite{ang13}.

The D1S and D1ST2a results are almost overlapping, with differences smaller than 0.5\%, 
and this indicates again that the tensor terms in the interaction do not produce noticeable 
effects on the charge radii.

We observe a systematic difference, of about $1\%$, 
between the results obtained with the D1S and the D1M interactions, the 
latter one producing smaller values. This difference improves the 
description of the data obtained with the D1M force with respect to the D1S and D1ST2a interactions. 
We remark that in the set of empirical information used to select the parameters 
of the D1M force, the charge radii values of 707 nuclei were considered \cite{gor09}. 
This is probably the reason of the better agreement with the experiment obtained with the D1M interaction.

In any case, it is very striking that all the IPM calculations fail to describe the charge radius of $^{48}$Ca, a doubly magic nucleus, and, in general, the charge radii of the nuclei between $^{40}$Ca and $^{48}$Ca. 
This is a well known problem since 1968 \cite{fro68}, and it has been shown to be present 
in different kinds of IPMs (see for example the results of Refs.\cite{abo92,wer96,nak19}). 

\begin{figure}[!b] 
\begin{center} 
\includegraphics [scale=0.4,angle=0]{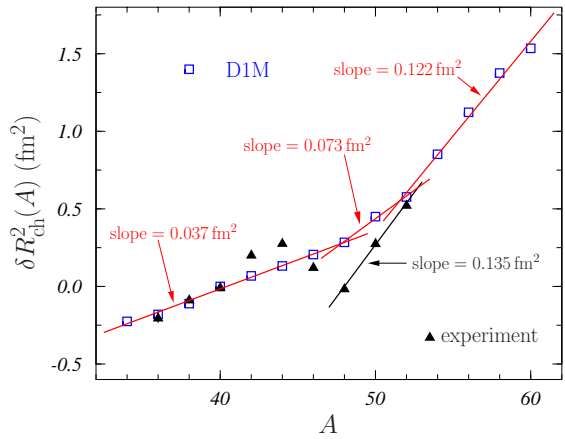} 
%\vskip -3.0 cm 
\caption{\small Isotope shifts, Eq. (\ref{eq:shift}) obtained in the IPM calculations 
with the D1M interaction  (open blue squares) and experimental results of Refs. \cite{gar16,mil19,woh81,pal84,ver92} 
(solid black dots). 
The linear fits to the three sections of the IPM results and to the experimental values for the heavier Ca isotopes 
are indicated by the straight lines. The angular coefficients, slopes, of these lines are also given.
}
\label{fig:slopes-IPM} 
\end{center} 
\end{figure} 

The behavior of the isotopes shifts that we have obtained with our IPM can be separated in three regions. As example of 
this analysis, we show again in Fig. \ref{fig:slopes-IPM} the comparison between the isotope shifts
obtained with the D1M interaction (blue open squares) and the experimental values (black solid triangles). 
These latter values are the average of those quoted by different experiments  \cite{gar16,mil19,woh81,pal84,ver92}.
The red lines of the figure are the results of linear fits of the D1M data for three regions: 
nuclei lighter than $^{48}$Ca,  from $^{48}$Ca to $^{52}$Ca, and nuclei heavier than $^{52}$Ca.
The values of the angular coefficients, the slopes, of these lines are also indicated in the figure. 
Similar results are obtained for the other two forces considered.

We identify three regions also in the case of the experimental data, 
but, in this case, the trend is remarkably different from that found for the IPM. 
The behaviour of the data for the isotopes lighter than $^{40}$Ca 
and heavier than $^{48}$Ca is linear.
The value of the slope of the line fitting this latter set of data is  very 
similar to that of the IPM results for nuclei heavier than
$^{54}$Ca. The real problem is related to the experimental 
data of the Ca isotopes from $^{40}$Ca to $^{48}$Ca. 
Contrary to the IPM results, these data present a maximum for the $^{44}$Ca 
nucleus, and a minimum for the $^{48}$Ca that, as we have already 
pointed out, shows a null isotope shift, indicating that its charge r.m.s. radius 
is essentially the same as that of $^{40}$Ca.

The inclusion of the correlations modifies the IPM charge distributions and, consequently, 
the values of the charge r.m.s. radii. The effects of the SRC are mainly concentrated in the nuclear
interior, therefore the IPM radii are only slightly modified.  
Eq.~(\ref{eq:radius}) clearly indicates that the values of the r.m.s. radii are obtained by integrating the  
densities weighted by a factor $r^4$, against a factor $r^2$ related to the normalization. 
For this reason, the radii are more sensitive to the changes of the nuclear surface than 
those of the interior.
The LRC, which consider the coupling of the IPM ground state with low-lying surface collective excitations,
enlarge the charge distributions and increase the values of the r.m.s radii of about 15\%. 

\begin{figure}[!b] 
\begin{center} 
\includegraphics [scale=0.4,angle=0]{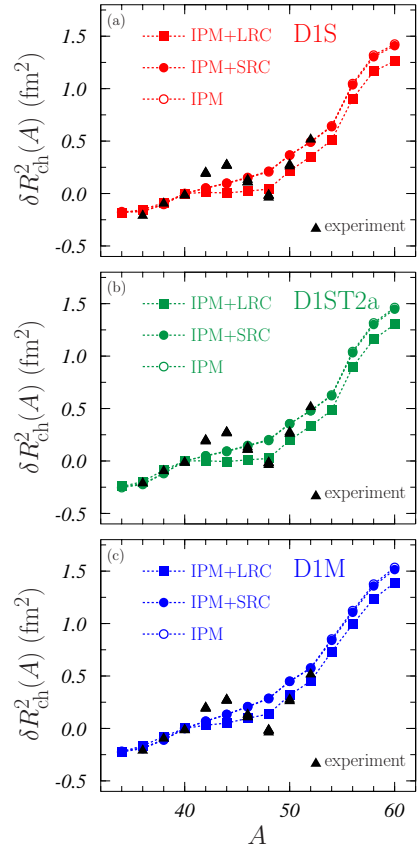} 
%\vskip -3.0 cm 
\caption{\small Effect of the short- and long-range correlations on the isotope shifts, defined in Eq. (\ref{eq:shift}). The IPM results (open circles) are compared to those found after adding SRC (solid circles) and LRC (solid squares). Results for the D1S, panel (a), D1ST2a panel (b) and D1M, panel (c), interactions are shown. Experimental data of Refs. \cite{gar16,mil19,woh81,pal84,ver92} are indicated by solid black triangles.
}
\label{fig:correl-effects} 
\end{center} 
\end{figure}

The effects of the correlations on the isotope shifts defined in Eq. (\ref{eq:shift}) are shown in Fig.~\ref{fig:correl-effects}.
The three panels present, separately, the results obtained for the three interactions considered. 
In each panel the open circles indicate the IPM results, the solid circles those obtained by considering the SRC, and the solid squares those including the LRC.

The effects of the SRC are extremely small, and the results obtained 
by taking them into account almost overlap with those of the IPM. 
On the contrary, the LRC produce a significant modification of the overall trend.

In the isotopes lighter than $^{40}$Ca, the LRC effects are negligible. However, already the IPM values
describe well the corresponding experimental data. 

The largest effects of the LRC are seen in the nuclei 
between $^{40}$Ca and $^{48}$Ca. The inclusion of the LRC generates a set of almost constant values for the D1S and D1ST2a interactions, while for the D1M interaction we found a moderate growth, though much smaller than that shown by the IPM results. The important point is that by including LRC a good description of 
the  $^{48}$Ca isotope shift for the D1S and D1ST2a interactions is obtained. 
The value found for the D1M interaction is not as good but, certainly, much better than that of the IPM. 
 
Above $^{48}$Ca, the trend of the isotope shift is roughly the same as in the IPM calculations, showing an enhancement with two well defined slopes for $A<52$ and $A>52$, respectively. For the nuclei heavier than this isotope, the growth rate of the isotope shifts is similar to the experimental one, as determined from the values of $^{48}$Ca, $^{50}$Ca and $^{52}$Ca.

\section{Discussion and conclusions}

In this work, we have presented the results of our study of the isotope shifts of even-even Ca isotopes. 
We first carried out IPM calculations, based on a HF+BCS approach, by using three finite-range 
effective interactions. The HF and BCS steps of our calculations have been executed by consistently
using the same interaction. The results obtained do not show any particular sensitivity to the interaction used, 
especially to the presence of tensor terms. These IPM results agree with the experimental values 
for the lighter Ca isotopes (with $A\leq 40$) and describe the growth rate of the heavier ones 
(with $A\geq 50$). The problem is the failure in describing the behavior of the experimental 
data of the nuclei from $^{40}$Ca to $^{48}$Ca, in particular the fact that these two doubly magic nuclei,
which in principle should be well described by the IPM, have the same charge radius. 
We did not find remarkable effects generated by the pairing, contrary to what has been claimed 
in Ref.~\cite{an20}.

We have extended our calculations by including short- and long-range correlations. 
The effects of the SRC, treated by using the model of Ref.~\cite{co95}, are irrelevant.
This result disagrees with the findings of Ref.~\cite{mil19a}. However, we found difficult to 
make a direct comparison between our and their definition of SRC 
and, therefore, to compare the specific contributions taken into account in both calculations. We remark that the fully microscopic calculation of Ref.~\cite{hag16} gives the correct proton radius of $^{48}$Ca. 

More significant are the effects of the LRC which we have described by extending the model 
of Ref.~\cite{ang01} in order to consider QRPA backward amplitudes. 
The LRC do not affect the behavior of the IPM in the region of the light and heavy Ca isotopes
where the IPM provides already a good description of the data. The results of the intermediate
region between $^{40}$Ca and $^{48}$Ca are strongly modified. The first, important, point is that
with the inclusion of LRC we obtain the same radii for the two doubly magic nuclei. The results
are better for the D1S and D1ST2a forces than for the D1M, but also in this latter case, the
improvement with respect to the IPM is evident. 

The second point is that, for this set of isotopes,
the LRC calculations generate almost constant values of the isotope shifts, contrary to the IPM 
results which show a continuous increase. Despite of a clear improvement, 
the trend of the experimental data, showing a maximum for $^{44}$Ca, is not yet 
well described. 

In Ref.~\cite{rei21} it has been pointed out the need of including terms related to 
proton and neutron magnetic moments and spin-orbit in order to obtain a precise description
of isotope shifts. These effects are very small, and, furthermore, 
they show a linear trend in the region between $^{40}$Ca and 
$^{48}$Ca, therefore they are unable to explain the behavior of the experimental results.

A good description of the data in the region of interest is provided by the
shell model calculation of Ref.~\cite{cau01}. By using our language, we may say that in this approach 
the LRC have been taken into account in a wider manner by including effects beyond the 
single quasi-particle excitations, which are the only ones considered in our QRPA calculations.

In conclusion, there is no problem in describing the new isotope shifts data measured for 
neutron rich Ca isotopes, and the data in the region between the two doubly magic isotopes
$^{40}$Ca and $^{48}$Ca indicate the relevance of the LRC. 

%%%%%%%%%%%%%%%%%%%%%%%%%%%%%%%%%%%%% 
\acknowledgments  
This work has been partially supported by the Junta de Andaluc\'{\i}a (FQM387), the Spanish Ministerio de Econom\'{\i}a y Competitividad (PID2019-104888GB-I00) and the European Regional Development Fund (ERDF). One of us (G.C.) thanks Sonia Bacca for useful discussions.


\begin{thebibliography}{10}
\expandafter\ifx\csname url\endcsname\relax
  \def\url#1{\texttt{#1}}\fi
\expandafter\ifx\csname urlprefix\endcsname\relax\def\urlprefix{URL }\fi
\expandafter\ifx\csname href\endcsname\relax
  \def\href#1#2{#2} \def\path#1{#1}\fi

\bibitem{gar16}
R.~F. {Garcia Ruiz} {\it et~al.,} Nat. \ Phys. {\bf 12}, 594 (2016).

\bibitem{mil19}
A.~J. Miller {\it et~al.,} Nat. \ Phys. {\bf 15}, 432 (2019).

\bibitem{tan20}
M.~Tanaka {\it  et~al.,} Phys. \ Rev. \ Lett. {\bf 124},  102501 (2020) .

\bibitem{fay00}
S.~A. Fayans, S.~V. Tolokonnikov, E.~L. Trykov and D.~Zawischa, Nucl. \ Phys. \ A
 {\bf  676}, 49 (2000).

\bibitem{klu09}
P.~Kl{\"u}pfel, P.-G. Reinhard, T.~J.B{\"u}rvenich and J.~A. Maruhn, Phys. Rev. C {\bf 79}, 034310 (2009). 

\bibitem{rei17}
P.-G. Reinhard and W.~Nazarewicz, Phys. \ Rev. \ C {\bf 95},  064328 (2017) .

\bibitem{pov93}
B.~Povh, K.~Rith, C.~Scholz and F.~Zetche, Teilchen und Kerne: Eine Einf{\"u}rung
  in die physicalischen Konzepte, Springer, Berlin, 1993.

\bibitem{ang14}
M.~Anguiano, A.~M. Lallena, G.~Co' and V.~{De Donno}, J. \ Phys. \ G {\bf 41}, 025102 (2014) .

\bibitem{ang15}
M.~Anguiano, A.~M. Lallena, G.~Co' and V.~{De Donno}, J. \ Phys. \ G {\bf 42},  079501 (2015).
.
\bibitem{edm57}
A.~R. Edmonds, Angular momentum in quantum mechanics, Princeton University
  Press, Princeton, 1957.

\bibitem{dec80}
J.~Decharg\'e and D.~Gogny, Phys. \ Rev. \ C {\bf 21}, 1568 (1980) .

\bibitem{ber91}
J.~F. Berger, M.~Girod and D.~Gogny, Comp. \ Phys. \ Commun. {\bf 63}, 365 (1991) .

\bibitem{gor09}
S.~Goriely, S.~Hilaire, M.~Girod and S.~P\'eru, Phys. \ Rev. \
  Lett. {\bf 102}, 242501 (2009) .

\bibitem{ang12}
M.~Anguiano, M.~Grasso, G.~Co', V.~{De Donno} and  A.~M. Lallena, Phys. \ Rev. \ C
  {\bf 86},  054302 (2012).

\bibitem{co95}
G.~Co', Nuovo \ Cimento \ A  {\bf 108}, 623 (1995).

\bibitem{wir95}
R.~B. Wiringa, V.~G.~J. Stoks and R.~Schiavilla, Phys.\ Rev. \ C {\bf 51}, 38 (1995).

\bibitem{ari07}
F.~{Arias de Saavedra}, C.~Bisconti, G.~Co' and A.~Fabrocini, Phys. \ Rep. {\bf 450}, 1 (2007).

\bibitem{row70}
D.~J. Rowe, Nuclear collective motion, Methuen, London, 1970.

\bibitem{len90}
H.~Lenske and J.~Wambach, Phys. \ Lett. B {\bf 249}, 377 (1990).

\bibitem{ang01}
M.~Anguiano and G.~Co', J. \ Phys. \ G {\bf 27}, 2109 (2001).

\bibitem{don17}
V.~{De Donno}, G.~Co', M.~Anguiano, and A.~M. Lallena, Phys. \ Rev. \ C {\bf 95}, 054329. (2017). 

\bibitem{cha07t}
F.~Chappert, Nouvelles param\'etrisation de l'interaction nucl\'eaire effective
  de {Gogny}, Ph.D. thesis, Universit\'e de Paris-Sud XI (France),
  http://tel.archives-ouvertes.fr/tel-001777379/en/ (2007).

\bibitem{bnlw}
{Brookhaven National Laboratory}, National nuclear data center.
  \url{http://www.nndc.bnl.gov/}.

\bibitem{woh81}
H.~D. Wohlfahrt, E.~B. Shera, M.~V. Hoehn, Y.~Yamazaki and R.~M. Steffen, Phys. \
  Rev. \ C {\bf 23}, 533 (1981).

\bibitem{pal84}
C.~W.~P. Palmer {\it et~al.,} J. \ Phys. \ B {\bf 17}, 2197 (1984).

\bibitem{ver92}
L.~Vermeeren, R.~E.~Silverans, P.~Lievens, A.~Klein, R.~Neugart, Ch.~Schulz, and F.~Buchinger,
Phys. \ Rev.\ Lett. {\bf 68}, 1679 (1992).

\bibitem{ang13}
I.~Angeli and K.~P. Marinova, At. \ Data \ Nucl. \ Data \ Tables {\bf 99}, 69 (2013).

\bibitem{fro68}
R.~F. Frosch {\it et~al.,} Phys. \ Rev. {\bf 174}, 1380 (1968).

\bibitem{abo92}
Y.~Aboussir, J.~M. Pearson, A.~Dutta and F.~Tondeur, Nucl. \ Phys. \ A {\bf 549}, 155 (1992).

\bibitem{wer96}
T.~R. Werner {\it et~al.,} Nucl. \ Phys. \ A {\bf 597}, 327 (1996).

\bibitem{nak19}
H.~Nakada, Phys. \ Rev. \ C {\bf 100}, 044310 (2019).

\bibitem{an20}
R.~An, L.-S. Geng and S.-S. Zhang, Phys. \ Rev. \ C {\bf 102}, 024307 (2020).

\bibitem{mil19a}
G.~A. Miller {\it et~al.,} Phys. \ Lett. \ B {\bf 793}, 360 (2019).

\bibitem{hag16}
  G. Hagen {\it et al.}, Nat. \ Phys. {\bf 12}, 186 (2016).

\bibitem{rei21}
P.-G. Reinhard and W.~Nazarewicz, Phys. \ Rev. \ C {\bf 103}, 054310 (2021) .

\bibitem{cau01}
E.~Caurier {\it et~al.,} Phys. \ Lett. \ B {\bf 522} (2021) 240.

\end{thebibliography}
\end{document}